\documentclass[aps,prl,superscriptaddress,twocolumn]{revtex4-1}

\usepackage{graphicx}
\usepackage{amsmath}
\usepackage{relsize}

\usepackage[colorlinks=true,citecolor=blue,linkcolor=magenta]{hyperref}
\hypersetup{
pdfauthor = {M. Batzer},
colorlinks = true, linkcolor = blue, urlcolor=blue, bookmarksnumbered =  true}

\begin{document}


\title{Single crystal diamond pyramids for applications in nanoscale quantum sensing}



\author{Marietta Batzer}
\affiliation{Department of Physics, Klingelbergstrasse 82, University of Basel}
\author{Brendan Shields}
\affiliation{Department of Physics, Klingelbergstrasse 82, University of Basel}
\author{Elke Neu}
\affiliation{ Universit\"at des Saarlandes, Fakultät NT, Physik, Campus E2.6, 66123 Saarbr\"ucken, Germany}
\author{Claudia Widmann}
\affiliation{Frauenhofer Institut f\"ur angewandte Festk\"orperphysik, Tullastrasse 72, 79108 Freiburg}
\author{Christian Giese}
\affiliation{Frauenhofer Institut f\"ur angewandte Festk\"orperphysik, Tullastrasse 72, 79108 Freiburg}
\author{Christoph Nebel}
\affiliation{Frauenhofer Institut f\"ur angewandte Festk\"orperphysik, Tullastrasse 72, 79108 Freiburg}
\author{Patrick Maletinsky }
\affiliation{Department of Physics, Klingelbergstrasse 82, University of Basel}



\date{\today}
\begin{abstract}
We present a new approach combining top down fabrication and bottom up overgrowth to create diamond photonic nanostructures in form of single-crystalline diamond nanopyramids. 
Our approach relies on diamond nanopillars, that are overgrown with single-crystalline diamond to form pyramidal structures oriented along crystal facets.
To characterize the photonic properties of the pyramids, color centers are created in a controlled way using ion implantation and annealing. We find very high collection efficiency from color centers close to the pyramid apex. We further show excellent smoothness and sharpness of our diamond pyramids with measured tip radii on the order of $10~$nm.
Our results offer interesting prospects for nanoscale quantum sensing using diamond color centers, where our diamond pyramids could be used as scanning probes for nanoscale imaging. There, our approach would offer significant advantages compared to the cone-shaped scanning probes which define the current state of the art.

\end{abstract}
\maketitle





Optically active point defects in solid-state hosts, also known as color-centers, form attractive, atom-like systems, offering vast opportunities in the field of quantum science and technology. 
Their spin states and optical transitions can be harnessed for applications ranging from quantum communication\,\cite{aharonovich_diamond-based_2011}, quantum networks\,\cite{Wehner2018a} to quantum sensing\,\cite{Rondin2014a,Degen2017a}.
The various color centers occurring in diamond have proven particularly relevant in this development and have already found applications in nearly all fields of quantum science and technology.
Nanoscale quantum sensors using individual, color center based electron spins in diamond have attracted particular interest, triggered by recent success in, e.g. nanoscale imaging of superconductors\,\cite{Thiel2016a} and ultrathin magnets\,\cite{Tetienne2014a,Thiel2019a}, as well as high-frequency probing of spin waves\,\cite{Du2017a}.

Such nanoscale quantum sensors live up to their full potential when employed in a scanning probe configuration using atomic force microscopy (AFM) tips decorated by single spins\,\cite{Rondin2014a}. 
This approach allows for precise, sub-nanometer positioning of the quantum sensor and thereby yields optimized resolution and sensitivity. 
While early implementations of this concept have focused on grafting color center containing nanodiamonds onto AFM tips\,\cite{Rondin2014a,Tetienne2014a,Kuhn2001a}, recent work increasingly focused on ``top-down'' fabrication of monolithic AFM tips from high-purity, single-crystalline diamond (SCD)\,\cite{Maletinsky2012a,Appel2016a}.
This approach combines several advantages: It yields highly robust tips, amenable to operation in harsh environments, such as cryogenic conditions\,\cite{Thiel2016a}. It mitigates optical blinking and excess spin dephasing which are both ubiquitous in nanocrystals\,\cite{Galland2012a}. And lastly, it allows for tailoring the photonic properties of the tips to yield high fluorescence collection efficiencies and thereby sensitivity\,\cite{Momenzadeh2015a}.

 \begin{figure}[t]
	\centering
	\includegraphics[width=\linewidth]{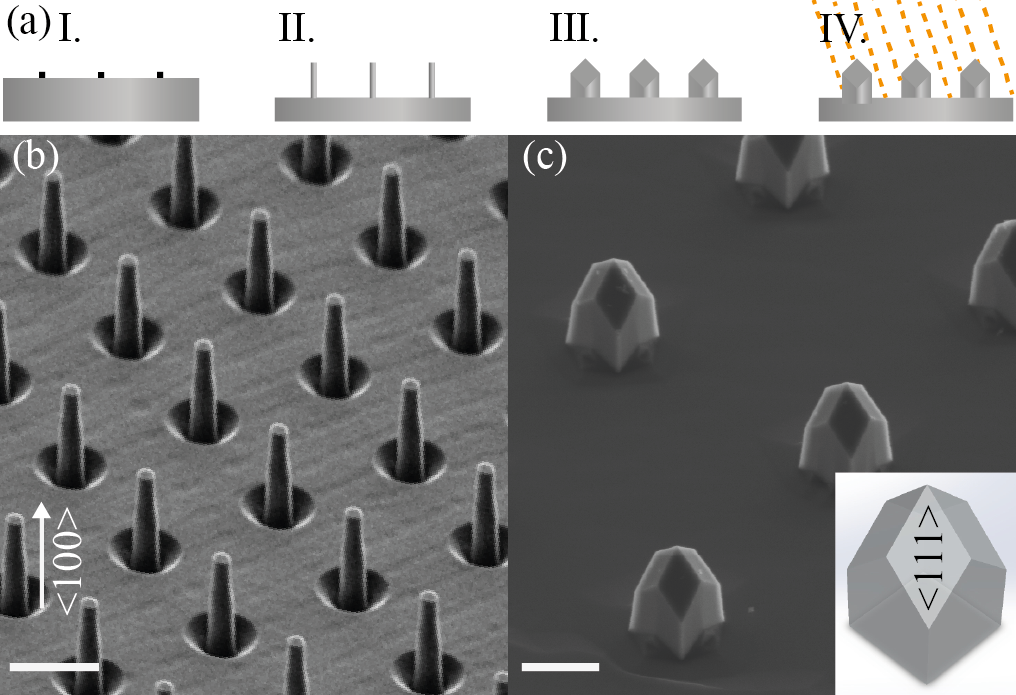}
	\caption{(a) Schematic of the nanofabrication and overgrowth processes to obtain single-crystal diamond (SCD) pyramids. I: Definition of etch-masks by e-beam lithography; II: Reactive ion etching of diamond nanopillars; III: Overgrowth of diamond material to form pyramids; IV: $^{14}$N-ion implantation and annealing. (b) Scanning electron microscopy (SEM) image of representative diamond nanopillars after nanofabrication (step II) and (b) after diamond overgrowth (step III). Scalebar $\widehat{=}~1~\mu$m in both cases. Both SEM images were recorded at a 45$ ^{\circ} $ tilt angle from the sample normal. 
	Inset: Schematic of the resulting pyramids with a $\left<111\right>$ crystal facet indicated.}
	\label{fig:fig1}
\end{figure}

However, the same approach also comes with several drawbacks. On the one hand, the hardness and chemical inertness of SCD requires harsh plasma etching procedures in the nanofabrication processes, which leaves the SCD surface damaged and results in non-ideal coherence properties of the sensing spins\,\cite{Myers2014a}. On the other hand, all single color center SCD scanning probes demonstrated up to now assume the approximate shape of a truncated cone, with a relatively blunt, circular end-facet of $\sim200~$nm diameter. While this shape has proven beneficial for the tips' photonic properties\,\cite{Momenzadeh2015a}, it is far from ideal for AFM performance for two reasons: First, the bluntness of these tips prevents simultaneous high-resolution AFM imaging with single spin magnetometry, which is relevant when imaging samples with non-planar geometries. Second, pillars with circular end-facets require excellent angular alignment to be in full contact with the sample, which typically results in increased spin-sample distances and a resulting loss in spatial resolution and sensitivity per source strength\,\cite{Rondin2014a}.

In this work, we present a new approach to realising all-diamond tips for scanning-probe, nanoscale quantum sensing, which has the potential to address all the drawbacks of previous approaches highlighted above. For this, we combine aspects of ``top-down'' (etching) and ``bottom-up'' (overgrowth) fabrication to yield nanometer-sharp, pyramidal diamond tips, as illustrated in Fig\,\,\ref{fig:fig1}(a). Specifically, we overgrow SCD nanopillars, fabricated via reactive ion etching, with high purity SCD by chemical vapor deposition (CVD).  The highly anisotropic CVD growth transforms these pillars into SCD pyramids\,\cite{nishibayashi_homoepitaxial_2000,jaffe_deterministic_2019}. The pyramids show high collection efficiencies for color center photo luminescence which together with their sharp apex radius of curvature of $\sim10~$nm makes them perfectly amenable for future applications as scanning probes.


To create the pillar template for overgrowth (Fig.\,\ref{fig:fig1}(a)), we fabricated $200~$nm-diameter, cylindrical diamond nanopillars with $\sim2~\mu$m length on a $\left<100\right>$-oriented electronic grade SCD substrate
grown in a custom designed ellipsoidal microwave plasma enhanced chemical vapor deposition reactor\,\cite{Widmann2016a}. Details of the nanofabrication process based on electron-beam lithography and reactive ion etching have been reported elsewhere\,\cite{Appel2016a,Widmann2017c}. 
The nanopillars were then overgrown with few $\mu$m of diamond by microwave plasma assisted chemical vapour deposition. Key growth parameters include a substrate temperature of $850^\circ$C, microwave power of $2.1~$kW, a chamber pressure of $200~$mbar and a gas mixture of $3:1$ [CH$_4$]:[O$_2$]  with a methane concentration of $0.3~\%$ in the process gas\,\cite{Widmann2017b}. 
Details of the growth process and growth apparatus can be found elsewhere\,\cite{Widmann2017b} and will be further discussed in a future publication.

As evidenced by the representative scanning electron microscopy (SEM) image shown in Fig.\,\ref{fig:fig1}(b), pillar overgrowth leads to the formation of pyramids with well defined geometry and excellent uniformity across the sample. Based on the fourfold symmetry, the orientation of the pyramids with respect to the $\left<100\right>$-oriented diamond surface and based on the angle between the top facets we identify these facets as $\{111\}$-planes of the diamond crystal, consistent with previous reports on diamond nanopillar overgrowth \,\cite{jaffe_deterministic_2019}. 
Remarkably, our structures exhibit near-perfect pyramidal shape. The anisotropy of diamond growth is conveniently characterized by the growth parameter $\alpha$\,\cite{bogatskiy_geometric_2015}, which quantifies the diamond growth rate along the $\{100\}$-planes, $V_{\{100\}}$, normalized to the growth rate $V_{\{111\}}$ along the $\{111\}$-planes, i.e. 
 \begin{equation}
\alpha = \sqrt{3}\frac{V_{\{100\}}}{V_{\{111\}}}.
\label{eq1}
\end{equation}
 Based on the observed pyramid shape, and using a recently established formalism and software\,\cite{bogatskiy_geometric_2015}, we estimate $\alpha\approx3.1$ for our diamond growth\,\cite{Gracio2010a}. The large $\alpha$ parameter results in the sharp pyramid apex we observe, in contrast to previous reports, where pyramids showed a characteristic truncation with a remaining $\{100\}$-facet\,\cite{jaffe_deterministic_2019}.

\begin{figure}
	\centering
	\includegraphics[width=\linewidth]{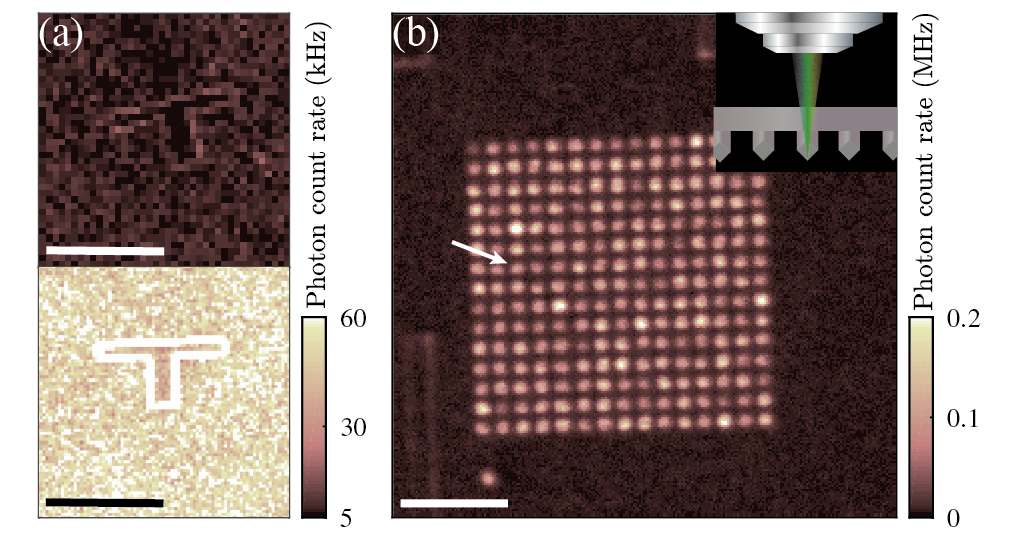}
	\caption{ (a) Confocal fluorescence map of the same part of an overgrown diamond substrate without pyramids before (top) and after (bottom) $^{14}$N implantation (the visible "T-shaped" structure is an alignment marker). (b) Confocal image of an array of overgrown pyramids with fluorescence collected for wavelengths below $700~$nm. The arrow indicates the pyramid which was further investigated in Fig.\,\ref{fig:fig3}(a)\&(b). The data were recorded at excitation powers of $150~\mu$W and $100~\mu$W in panels (a) and (b), respectively. The scale bars in all images is $10~\mu$m. Inset: Schematic (not to scale) of the sample orientation during measurement. The diamond sample thickness was $\sim20~\mu$m.}
	\label{fig:fig2}
\end{figure}

To investigate the photoluminescence (PL) emitted from color centers inside the pyramids, we employed a home-built confocal microscope\,\cite{Appel2016a} with numerical aperture NA$~0.8$ and continuous laser excitation at $532~$nm. 
Prior to confocal characterisation, the sample was boiled in a $1:1:1$ mixture of nitric, sulfuric, and perchloric acids to remove surface residues and ensure oxygen termination of the diamond surface. 
Inspection of the as-grown samples treated as described before showed no significant fluorescence, indicating the high purity of the overgrown as well as substrate material. 
After implantation (Innovion; $^{14}$N fluence: $3$e$11~$ions/cm$^2$, energy: $12~$keV, sample tilt: $7^\circ$), annealing ($4~$h at $400^{\circ}$C, $10~$h at $800^{\circ}$C and $4~$h at $1200^{\circ}$C) and a second acid treatment, we observe significant color-center fluorescence from the diamond surface (Fig.\,\ref{fig:fig2}(a)), 
containing spectral signatures of both negatively charged Nitrogen-Vacancy (NV$^-$) and Silicon-Vacancy (SiV$^-$) centers, 
the latter of which result from Si impurities introduced during overgrowth.  
Based on numerical modeling of the implantation process (SRIM software package run with lattice binding energy $1.5~$eV, surface binding energy $4.5~$eV and displacement energy $45~$eV\,\cite{Uzansaguy1995a,Ma1999a,Reinke1996a}) we expect the observed color centers to be located within $\sim20~$nm from the diamond surface\,\cite{Favaro2017a,Vandam2019a}. 

\begin{figure}[t!]
	\centering
	\includegraphics[width=\linewidth]{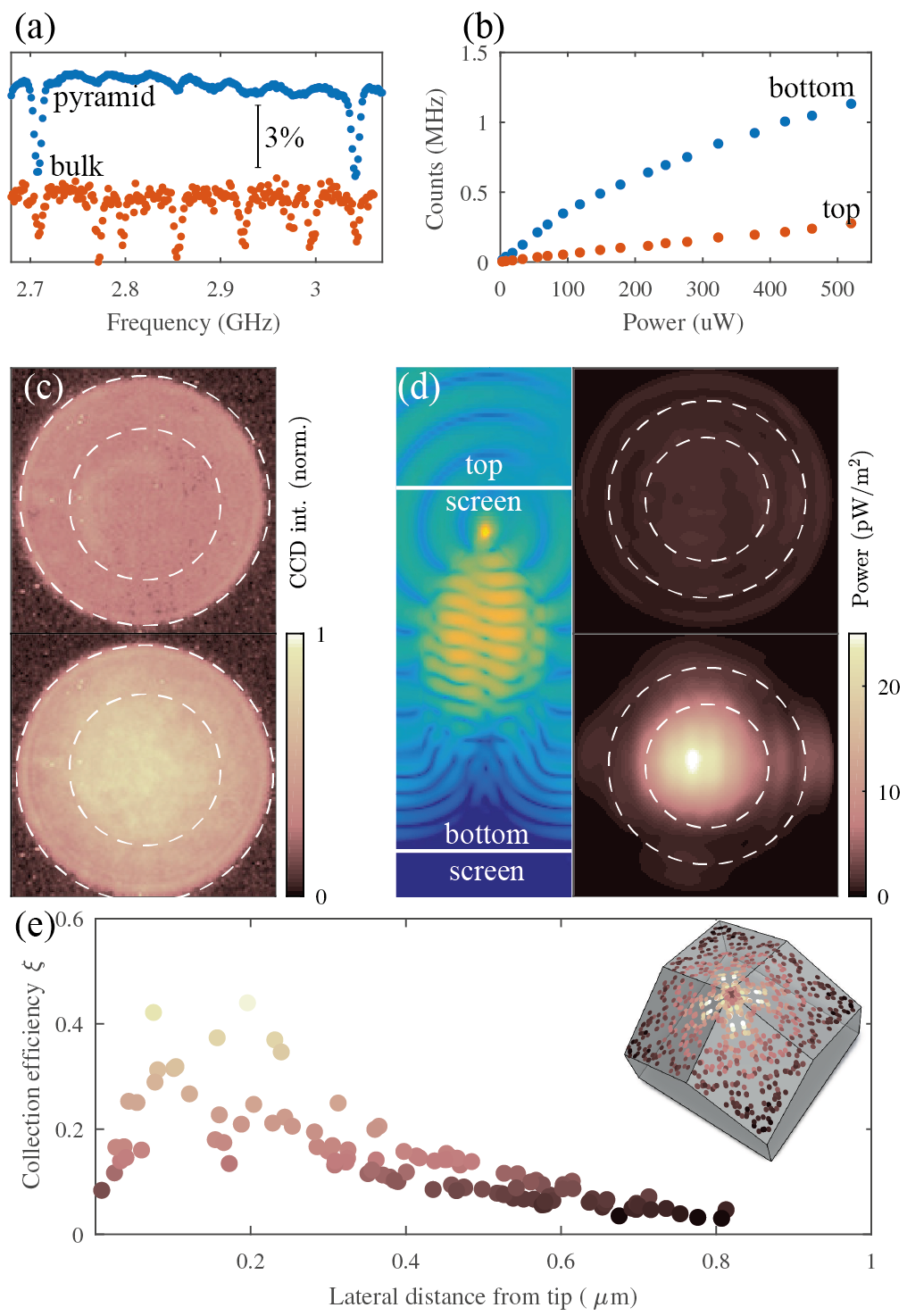}
	\caption{ (a) Representative optically detected electron spin resonance (ODMR) from a diamond nanopyramid (highlighted in Fig.\,\ref{fig:fig2}(b)) and from surrounding bulk. The single pair of ODMR lines indicates emission from a single or few NVs in the pyramid. 
	(b) Fluorescence saturation curves collected from the top (red) and bottom (blue) side of the same diamond pyramid.
	(c) Experimentally measured back focal plane images of emission from a representative pyramid for top and bottom collection (top and bottom, respectively) 
	(d) Finite difference time domain (FDTD) simulation of fluorescence emission from a point dipole in the overgrown diamond pyramid. 
	Left: cross-section through the diamond pyramid with the emitter placed at the pyramid apex. 
	Right: Simulation of far-field power distribution patterns collected from the pyramid top and bottom (top and bottom row, respectively).
	For all panels, white circles correspond to numerical apertures (NAs) of $0.8$ and $0.5$.
	(e) Collection efficiency for NA $0.8$ through the pyramid bottom ($\xi^{\rm bottom}$), as a function of lateral NV distance from the pyramid apex. The inset shows the NV locations considered, where the color of the points encodes the value of $\xi^{\rm bottom}$.}
	\label{fig:fig3}
\end{figure}

For a detailed characterisation of the diamond pyramids' photonic properties, we focused on NV$^-$ emission and suppressed unwanted signals from SiV$^-$ emission by an appropriate short-pass filter (Thorlabs FES0700, Cut-Off wavelength: 700 nm).
A resulting confocal map recorded through the pyramid-base (the fluorescence collection direction relevant for the scanning probe application we target in the future) is shown in Fig.\,\ref{fig:fig2}(b).
NV$^-$ emission from the pyramids is significantly brighter than NV$^-$ fluorescence from the nearby unstructured diamond, which already indicates efficient waveguiding and enhanced collection efficiencies from these structures\,\cite{nelz_color_2016} (another potential explanation, namely the preferential incorporation of Nitrogen into $\{111\}$-facets during diamond growth\,\cite{Samlenski1996a}, can be excluded based on the low density of residual Nitrogen ($<1~$ppb\,\cite{Widmann2017a}) in the overgrown material.

The NV$^-$ fluorescence both from the unstructured surface and from the pyramids exhibits clear signatures of optically detected magnetic resonance (ODMR)\,\cite{Gruber1997}. 
Figure\,\ref{fig:fig3}(a) shows two representative ODMR traces, one from a pyramid and one from the surrounding unstructured surface, which were obtained in the same bias magnetic field applied in a direction not aligned with any of the diamond samples' symmetry axes to distinguish different NV families. 
Interestingly, the majority of pyramids investigated show a single pair of ODMR lines, whereas the NV$^-$ ensemble observed on the unstructured part of the diamond (Fig.\,\ref{fig:fig3}(a)) as expected shows four such pairs, where each pair corresponds to an NV family aligned along the same, 111-equivalent crystal direction.

A statistical analysis of the number of ODMR lines observed in the pyramids allows for an estimation of the NV density in these structures. Out of a total of seven investigated pyramids, three showed a single pair of well resolved ODMR lines, while the remaining pyramids had either two such pairs or unclear ODMR traces. Assuming a Poissonian distribution of the number of NV's per pyramid, the measured probability of $3/7$ of observing a single ODMR line yields an expectation value of $\sim 2.6$ NVs per pillar. This estimation deviates by more than a factor of ten from our expected NV density. Specifically, using either our implantation parameters and the known N$\rightarrow$NV$^-$ conversion efficiency\,\cite{Pezzagna2010} or, alternatively and consistently, the brightness of NV$^-$ fluorescence from the unstructured surface, compared to a well-known reference sample, we estimate an NV$^-$ density of  $\sim 30~\mu$m$^{-2}$ in the sample under investigation. For the given pyramid footprint, this would result in an average number of 45 NV$^-$ centers per pyramid in stark contrast to our above estimation. As we will show in the following, this discrepancy results from the nanophotonic properties of the pyramids: Optical waveguiding is most efficient for NV centers in the vicinity of the pyramid apex, and as a result, our experiment most efficiently detects NV fluorescence from this subset of NVs only.

In order to experimentally assess the photonic properties of the nanopyramids, we measured the angular emission patterns from their embedded NV$^-$ centers by back focal plane (BFP) imaging.
Figure\,\ref{fig:fig3}(c) top and bottom show, respectively, the resulting BFP emission patterns for NV$^-$ collection through the pyramid apex (further referred to as ``top-collection'') and through the pyramid base and substrate (``bottom-collection'').
In these subfigures and the following, white circles indicate collection NA's of $0.5$ and $0.8$ (the NA of our collection optics).
While top collection shows diffuse emission into the whole collection NA, bottom-collection shows a BFP emission patterns which is more centred on the optical axis and shows emission predominantly within a collection NA of $\sim0.5$. 
NV$^-$ emission from the nanopyramids shows significant directionality: Approximately five times more fluorescence is emitted to the bottom compared to the top side, as evidenced by both the signal integrated over the BFP images and fluorescence saturation curves collected from the top and bottom sides for the same pyramid (Fig.\,\ref{fig:fig3}(b)). 
This observed directionality of color center emission from nanopyramids is in qualitative agreement with previous reports\,\cite{choi_enhancing_nodate,nelz_color_2016}.

For a better understanding of the observed BFP emission patterns, we performed numerical simulations using the finite-difference time-domain (FDTD) module of the commercially available software Lumerical. 
There, we considered individual optical dipoles emulating NV$^-$ emission and calculated the far-field emission patterns corresponding to our top- and bottom-collection BFPs (Fig.\,\ref{fig:fig3}(b), top and bottom row, respectively).
The NV locations were randomly chosen on one of the top facets of the pyramid at a depth of $20~$nm below the diamond surface.
To each location, we randomly assigned one of the four possible NV orientations and performed our calculations for two orthogonal optical dipoles lying in the plane orthogonal to the NV direction.

The simulations qualitatively reproduce our experimental BFP images and show a clear tendency of waveguiding of NV$^-$ emission towards the pyramid bottom. To quantify this directionality we consider the commonly used collection factor\,\cite{Fuchs2018a}
    \begin{equation}
   \xi= \frac{\Gamma_{\rm NA}}{\Gamma_{\rm rad}},
   \label{eq2}
   \end{equation}
where $\Gamma_{\rm rad}$ is the radiative emitter decay rate in a homogeneous medium and $\Gamma_{\rm NA}$ is the rate of far field photons emitted into the collection NA (here with NA$=0.8$).
Our simulations for the case of  NV$^-$ concentration near the pyramid apex yield a top and bottom-side collection factor $\xi^{\rm bottom}$= 0.20 and $\xi^{\rm top}$=0.058, whose ratio $\xi^{\rm bottom}/\xi^{\rm top}=3.48$ is in reasonable agreement with our experimental finding of $\xi^{\rm bottom}/\xi^{\rm top}\sim 3.2$ and previous reports on similar structures\,\cite{nelz_color_2016,jaffe_deterministic_2019}.

Our simulations also shed light on our observed discrepancy between the NV$^-$ density estimated from the optical (ODMR) signal from the pyramids as compared to the estimated NV$^-$ density based on implantation parameters (Fig. 3(a)). Specifically, we find that $\xi^{\rm bottom}$ is strongly dependent on the lateral NV$^-$ location within the pyramid. Figure\,\ref{fig:fig3}(e) shows $\xi^{\rm bottom}$ as a function of this position, which we parametrize by the radial distance $r$ of the NV$^-$ location from the pyramid center. The simulation was performed for a selection of $110$ randomly placed NVs, where NVs with same $r$ but different  $\xi^{\rm bottom}$ correspond to different azimuthal positions of the NVs (see inset to Fig.\,\ref{fig:fig3}(e))  and/or different dipole orientations assigned to same NV position. 
The simulation shows that the NVs with the highest $\xi^{\rm bottom}$ are all located close to the pyramid apex, with $\xi^{\rm bottom}$ peaking at $\xi^{\rm bottom}_{\rm max}\sim0.5$ for $r\sim150~$nm. 
For a qualitative estimate, we postulate that the bottom collected NV signal be dominated by those NVs with $\xi^{\rm bottom}>\xi^{\rm bottom}_{\rm max}/2$; 
from our previous NV density density estimate ($\sim 30~\mu$m$^{-2}$), we then conclude that the emission from $\sim3$ NVs dominates the PL collected from a nanopyramid.

We stress that further factors might add to this qualitative argument to explain the lower NV density we observe in the nanopyramids compared to their surrounding. These include potential distortions of electric field lines by the nanopyramids during the $^{14}$N implantation process, which could lead to non-uniform implantation profiles, or different degrees of incorporation of native nitrogen or vacancies during growth of the pyramids compared the the surrounding bulk. 

 \begin{figure}
 	\centering
 	\includegraphics[width=\linewidth]{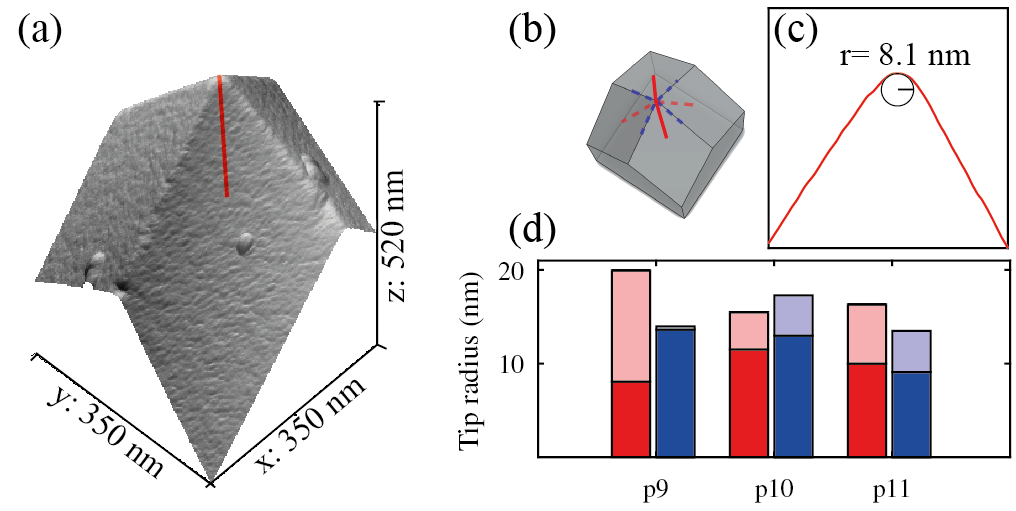}
 	\caption{(a) AFM image of pyramid 9 (p9). (b) Schematic sketch of a pyramid with 
 		colored lines indicating the positioning of the line cuts (blue: along the edge or red: across the facet) used to fit the tip radii in (d). (c) line cut through the AFM structure highlighted in (a), with osculating circle fitted to the tip (black) (d) Histogram of tip radii for three different pyramids. The darker (light) bar indicates the minimal (maximal) radii measured on each pyramid whereas colors symbolize the direction of the line cut. }
 	\label{fig:fig4}
\end{figure}

One of the key benefits of working with bottom up structures is the improved surface quality of CVD grown diamond compared to diamond etched in a plasma. To characterize and quantify the surface and the sharpness of our diamond pyramid tips, we performed detailed AFM imaging on representative pyramids from our sample  (Fig.\,\ref{fig:fig4}(a)).
The AFM image reflects the clear pyramidal shape of the tip already seen in our SEM investigation and shows flat crystal facets.  
The visible surface roughness with root-mean-square amplitude $\sim2~$nm results from a thin metal coating (Ti, $2~$nm) applied to the sample to prevent charging during SEM imaging. 
Based on previous studies of diamond growth under similar conditions\,\cite{Widmann2017a}, we estimate the roughness of the diamond pyramids to be $<1~$nm.
More importantly though, our AFM image confirms the sharpness of the pyramid tips already visible in Fig.\,\ref{fig:fig1}(b).
To quantify the tip sharpness, we extracted line cuts across pyramid facets and edges as illustrated in Fig.\,\ref{fig:fig4}(b) and fitted osculating circles to these line profiles (Fig.\,\ref{fig:fig4}(c)).
 Figure\,\ref{fig:fig4}(d) summarizes our findings and shows the minimal and maximal radii found in each direction for a selection of three pyramids. Importantly, all tip radii found were in the range $8~$nm$...20~$nm, which on one hand demonstrates the remarkable sharpness of these tips and on the other hand suggests that the values determined here are not masked by the radius of the AFM tip employed for imaging.

With our work we have established diamond pyramids created by CVD overgrowth of diamond nanopillars as an attractive avenue for future all-diamond scanning probe quantum sensors. We demonstrated a scalable process that yields sharp diamond pyramids which appear highly attractive as robust, high-resolution AFM tips in general and for nanoscale quantum sensing in particular. In the latter case, the sharpness of the tip would ensure close proximity of a color-center quantum sensor placed at the apex of the pyramid to a sample of interest. An improvement over existing approaches\,\cite{Maletinsky2012a}, which could promote spatial resolution in such imaging to the sub-$10~$nm range. Furthermore, we demonstrated that the pyramids are effective photonic nanostructures, which yield high-efficiency fluorescence collection, on par with currently available nanoscale quantum sensing technologies\,\cite{Maletinsky2012a,Zhou2017a}. This, together with the native diamond surface, which supports long spin coherence times for near-surface color center spins\,\cite{Kato2017a}, will help further improve sensitivities in, e.g. scanning NV magnetometry. 

Two key steps still need to be addressed in order to employ such diamond pyramids for scanning probe quantum sensing experiments: The overgrowth method demonstrated here has to be combined with scanning probe fabrication and individual color centers need to be controllably created at the pyramid apex. The first requirement can easily be met, since scanning probe fabrication procedures are readily applicable to the diamond pyramids realized here. Color-center creation at the pyramid apex appears more challenging but could be achieved by ion implantation with nanoscale resolution, either by nano-implantation through AFM tips\,\cite{Riedrich-Moeller2015a}, or through focussed-ion beam implantation of colour centers\,\cite{Schroder2017a}. 
Our analysis of the photonic properties of the pyramids also suggests and alternative, scalable route to the same end: Namely to pursue the procedure outlined in this paper for scanning probe fabrication and to exploit the highly position-dependant collection-efficiency for NV PL to post-select for NVs with highest fluorescence collection efficiency, which would therefore be located in reasonable proximity to the pyramid apex.

\section{acknowledgments}

We thank U. Pieles (FHNW) for constant support throughout the thesis of M. Batzer. We gratefully acknowledge financial support through the EU Quantum Flagship project ASTERIQS (Grant No. 820394), through the NCCR QSIT (Grant No. 185902), the Swiss Nanoscience Institute, and through the Swiss NSF Project Grant No. 169321.

\bibliography{literature.bib}

\end{document}